\documentclass[11pt]{article}         
\usepackage{graphicx}
\usepackage{amsmath,amsthm,amsfonts,amssymb,latexsym}
\usepackage{eucal}
\usepackage{tikz}
\usepackage{everyshi}
\usepackage{xcolor}

\newtheorem{theorem}{Theorem}

\newtheorem{definition}{Definition}
\newtheorem{axiom}{Axiom}

\definecolor{darkgreen}{RGB}{0, 170, 0}

\begin{document}

\title{Electroweak Quantum Numbers\\in the $D_4$ Root System}

\author{Henrik Jansson\\henjansson@gmail.com}

\maketitle

\begin{abstract}
\noindent We introduce four fundamental quantum numbers based on the $D_4$ root system, giving a unified description of quarks and leptons. These numbers will make it possible to define electric charge in a simple way. By postulating a fundamental symmetry given by the binary tetrahedral group acting on this root system, we can easily prove conservation of electric charge. Also, possible electroweak interactions are given by a simple rule where the 24-cell acts on itself by quaternion multiplication. The three generations of fermions in the Standard Model are hereby identified with the three imaginary dimensions of the quaternions. 
\end{abstract}

\section{Introduction}\label{sec1}

During the 1970s, the Standard Model of particle physics became a cornerstone of modern physics, including the electromagnetic, weak and strong interactions of elementary particles. While the Standard Model has seen some remarkable success, it still suffers from several weaknesses, except being unable to incorporate gravity. One of its  shortcomings is its rather \textit{ad hoc} mathematical structure, giving the impression of a patchwork. This might be very reasonable from a historical point of view, where bits and pieces have been added to the theory alongside new experimental discoveries, but nevertheless it leaves a feeling of dissatisfaction and a lack of understanding of nature at its most fundamental level. It is the aim of this paper to investigate some ideas related to electroweak interactions, in order to possibly find a simpler and more elegant mathematical formulation of elementary particles and their interaction.

In this attempt there are three mathematical structures which play a pivotal role: the binary tetrahedral group $T_{24}$ (in some papers denoted by $T'$), the root system $D_4$ corresponding to the Lie algebra $\mathfrak{so}(8)$ and the Lie group Spin(8), and finally the quaternions, $\mathbb{H}$, which will be seen to act as a common arena for the other two structures. Let us take a closer look at each and one of them and some of their possible connections to physics seen thus far.

Already in the 1950s, finite symmetry groups were considered in connection with electric charge of particles, where $T_{24}$ was one of the groups under investigation \cite{Case}. Some twenty years later, discrete symmetries were investigated in the realm of quark masses \cite{Wilczek}. A more modern treatment of finite groups, including $T_{24}$, connected to flavor and fermion masses can be found in \cite{Frampton94}. The last few decades an interest in finite symmetry groups in particle physics has arisen from neutrino research \cite{Aranda, Feruglio, Eby, King}. In a recent paper \cite{Frampton23}, it is argued that the binary tetrahedral group ``has provided the most successful flavour symmetry in understanding simultaneously the three mixing angles both for quarks in the CKM matrix and for neutrinos in the PMNS matrix.'' An interesting perspective from a more mathematical point of view can be found in \cite{Wilson20, Wilson211, Wilson212, Wilson21}, where some of the ideas in \cite{Wilson21} will be touched upon in this paper.

Ever since Heisenberg's attempt to unify protons and neutrons into different states of a nucleon, Lie groups and their representations have been a crucial part of physics \cite{Huerta, Griffiths}. Even though Heisenberg did not succeed with this endeavour, the group under consideration, SU(2), has found its use in several other situations: spin $1/2$ systems in quantum mechanics, non-abelian gauge theories \cite{Yang}, and as the weak isospin part of the Standard Model gauge group $G_{\text{SM}}=$ U(1)$\times$SU(2)$\times$SU(3). In the 1970s, physicists began to use other Lie groups in Grand Unified Theories (GUTs), where the first attempts were based on SU(5) and Spin(10) \cite{Huerta, Ross, Griffiths}, as well as Spin(4)$\times$Spin(6) \cite{Huerta, Pati}. One group with a very special property, Spin(8), has in this context been ruled out because of its lack of complex representations, something considered a necessity to describe chirality \cite{Ross}\footnote{One exception is a paper from 1986 where an attempt is made to relate masses of elementary particles to the structure of Spin(8) and the octonions \cite{Smith}.}. However, its unique property, so-called \textit{triality} (a symmetry between the three eight dimensional irreps of Spin(8) \cite{Yokota}), might be connected to other kinds of symmetries within particle physics, and is closely related to triality of normed division algebras over $\mathbb{R}$, of which there only exists four: $\mathbb{R}$, $\mathbb{C}$, $\mathbb{H}$, and last, but not the least, the octonions, $\mathbb{O}$ \cite{Furey}. In the present paper, we claim that Spin(8), or rather its root system $D_4$, when recast in the language of quaternions, does have a key role to play in the particle puzzle found in nature.

While physics is unimaginable without the real and complex numbers, quaternions and octonions have never really been accepted as essential building blocks for physical theories. Nevertheless, some natural connections do exist between quaternions and relativity as well as quantum mechanics \cite{Conway12, Conway48, Dirac, Lambek}. (Even though the use of octonions in physics has been under consideration since at least 1973 \cite{Gursey}, it is still of a more speculative nature, but a lot of work is under way to possibly find some answers \cite{Furey, Furey2, Furey3, Furey4}). As mentioned above, the group SU(2) is crucial for the theory of the weak interaction. Now SU(2) $\cong S^3$ can be considered as the group of unit quaternions, which in turn contains the binary tetrahedral group (see section \ref{crossgen} for details). Hence, it might be reasonable that a finite subgroup such as $T_{24}$ has a role to play in a theory of electroweak quantum numbers. Also, in the theory of spin $1/2$ particles, of which the weakly interacting fermions are prominent examples, the Pauli spin matrices can (up to a factor of $i$) be identified with the basic quaternions $i, j$ and $k$. In this paper, the three imaginary dimensions of quaternions spanned by $i$, $j$ and $k$ will correspond to the three generations of fermions in the Standard Model. So after all, quaternions might be the trick of the trade for dealing with these aspects of particle physics.

\section{The 28-fold Way}\label{sec3}

\begin{figure}[!t]
  \centering
  \includegraphics[width=\linewidth]{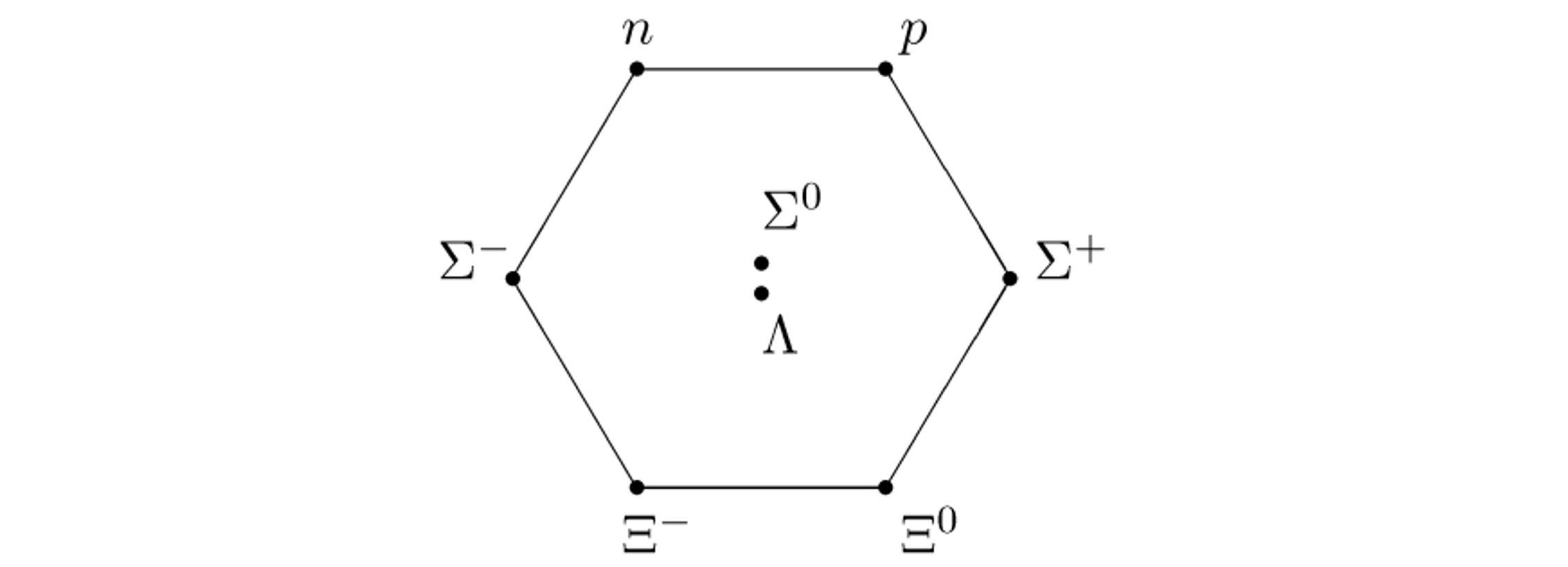}
\caption{The adjoint representation of $\mathfrak{su}(3)$ containing baryons.}
\label{fig:baryonoctet}
\end{figure}

In the 1950s, quarks were still not theorized or indicated by experiments. Instead, particle physicists had a huge number of potentially elementary particles to deal with. In an attempt to bring some order to that chaos, in 1961 Murray Gell-Mann (and independently Yuval Ne'eman) invented the so-called \textit{Eightfold Way} \cite{Gell-Mann61}, an approximate mass classification of baryons and mesons using irreps (irreducible representations) of the eight-dimensional Lie algebra $\mathfrak{su}(3)$. Since $\mathfrak{su}(3)$ has rank 2, the weight diagrams are two-dimensional and can easily be visualized. One example is the baryon octet illustrated in Fig. \ref{fig:baryonoctet} where the proton and neutron are found together with some other baryons of approximately the same mass.

Since the Eightfold Way was quite successful at the time of its invention, we think that a modern analogy could be beneficial. The structure of the Eightfold Way was explained by the introduction of the three quarks $u$, $d$ and $s$, corresponding to the three dimensional irrep of $\mathfrak{su}(3)$. However, there are three more quarks, $c$, $t$ and $b$, as well as six leptons and their anti-particles. It would be gratifying if all 24 elementary fermions (12 particle/anti-particle pairs) could be put in the same irrep of some Lie algebra. In this section, such a \textit{28-fold Way} of elementary particles will be suggested, but to motivate the ideas of this new way of organizing particles, we first want to spend some time on a simpler case. This will not have any direct physical relevance, but is included for illustrative purposes.

\subsection{Warm up} 
\begin{figure}[!t]
  \centering
  \includegraphics[width=0.5\linewidth]{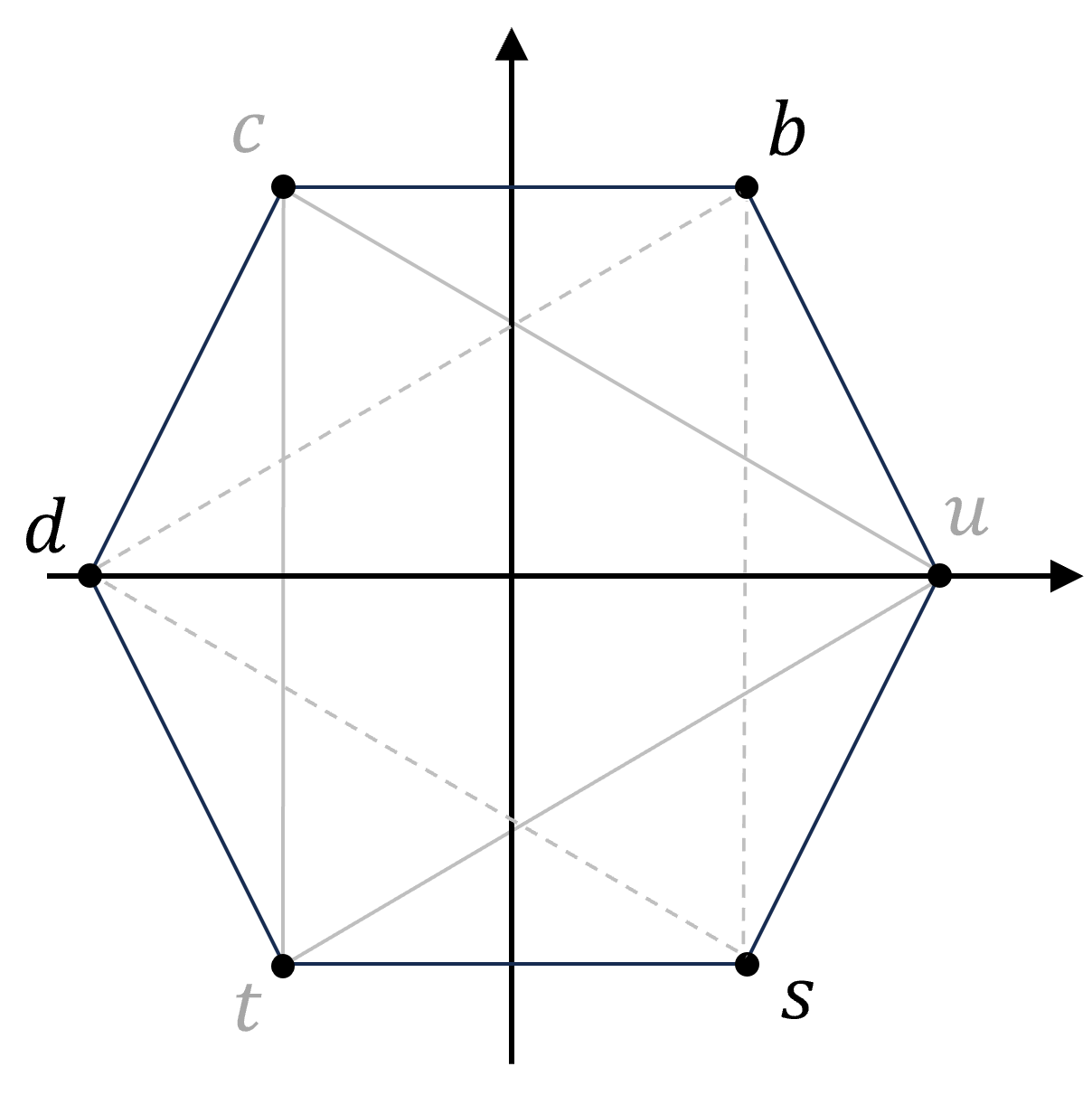}
\caption{All six quarks as roots of $\mathfrak{su}(3)$. There is a $Z_2$ symmetry within generations, and a $Z_3$ symmetry across generations. $Z_2 \times Z_3$ acts on the root system.}
\label{fig:quarkdiagram}
\end{figure}

Let us identify the six quarks with the roots of $\mathfrak{su}(3)$, see Fig. \ref{fig:quarkdiagram}. Since these roots are two-dimensional, we can treat them as complex numbers, i.e. $b = e^{\pi i/3} \in \mathbb{C}$ and so on. This makes it possible to transform one quark into another by simple operations on complex numbers. First, to change quark flavor within one generation, we just have to multiply by $-1 \in \mathbb{C}$ -- this operation generates a group of order two, $Z_2$. Second, to change flavor across generations, we rotate by $2\pi/3$. This can be accomplished by multiplication by $\omega = e^{2\pi i/3}$, e.g. $s = \omega d$. The rotation $\omega$ generates a cyclic group of order 3, $Z_3$. Hence, the group $Z_2 \times Z_3$ of order 6 acts on the hexagonal root system.

To include the leptons, we have to use another copy of this root system. Again the groups $Z_2$ and $Z_3$ will give the symmetries within and across generations. Alternatively, we could include three quarks/leptons and their anti-particles -- in this case $Z_2$ transforms a particle into its anti-particle.

\subsection{A classification scheme for quarks and leptons}

Even though we have now included all elementary fermions in the Standard Model, we would like to unify them into a single irreducible representation instead of using four copies of (the adjoint representation of) $\mathfrak{su}(3)$. This will be done by substituting $\mathfrak{su}(3)$ with the 28 dimensional $\mathfrak{so}(8)$ of rank 4, and using the quaternions $\mathbb{H}$ instead of the complex numbers $\mathbb{C}$. At the same time the group $Z_2$ will be replaced by the quaternion group $Q_8 = \{\pm 1, \pm i, \pm j, \pm k\}$ -- its meaning in this context will be explained later.

The root system of $\mathfrak{so}(8)$ consists of twelve pairs of roots:
\begin{eqnarray*}
\pm (1, \pm 1, 0, 0),\ \ \pm (1, 0, \pm 1, 0),\ \ \pm (1, 0, 0, \pm 1), & &\\
\pm (0, 1, \pm 1, 0),\ \ \pm (0, 1, 0, \pm 1),\ \ \pm (0, 0, 1, \pm 1), & &
\end{eqnarray*}
with all possible sign combinations. Hence, there will be exactly one root for each type of elementary fermion in the Standard Model. Noting that these roots are four-dimensional, we will identify them with elements of $\mathbb{H}$, so that $(1, 1, 0, 0) = 1 + i$, $(1, 0, -1, 0) = 1 - j$ etc. Since fermion types are usually referred to as ``flavors'' (``electroweak charge'' might have been a better name), the following definition seems appropriate.
\\
\begin{definition}
With a \textbf{(fermion) flavor} we mean a root of  $\mathfrak{so}(8)$.
\end{definition}
\mbox{}\\
\noindent So how are the flavors assigned to the different particles? Because of the high degree of symmetry, it is somewhat arbitrary where we start -- the choice for the rest of the particles will then follow. We will make the choice given in Table \ref{table:flavors}, where the positively charged particles will have $1$ in all non-zero entries, e.g. $u = (1, 1, 0, 0)$ and $e^+ = (0, 1, 1, 0)$. The flavor of an anti-particle will be the negative of the corresponding particle's flavor, i.e. $\bar{u} = (-1, -1, 0, 0)$\footnote{Note here that the bar does \textit{not} denote quaternion conjugation.}, $e^- = (0, -1, -1, 0)$ and so on. In the last column of Table \ref{table:flavors}, the subspace of $\mathbb{H}$ where each generation resides is included. ($\mathbb{R}_ {1ij}$ is shorthand for span$_\mathbb{R}\{1, i, j\}$ etc.)

\begin{table}[!t]
\centering
  \caption{Fermion types (flavors) as roots of $\mathfrak{so}(8)$, organized in generations.}
  \label{table:flavors}
  \vspace{0.5cm}
  \begin{tabular}{c  c  c}
     \hline
     \textbf{Fermions} & \textbf{Flavors} & \textbf{Subspace} \\
     \hline
     $d$, $u$, $\nu_e$, $e^+$ & $(1, \pm 1, 0, 0)$,  $(0, 1, \pm 1, 0)$ & $\mathbb{R}_{1ij}$ \\
    $s$, $c$, $\nu_\mu$, $\mu^+$  & $(1, 0, \pm 1, 0)$, $(0, 0, 1, \pm 1)$ & $\mathbb{R}_{1jk}$ \\
    $b$, $t$,  $\nu_\tau$, $\tau^+$ & $(1, 0, 0, \pm 1)$, $(0, \pm 1, 0, 1)$ & $\mathbb{R}_{1ki}$ \\
    \hline
  \end{tabular}
\end{table}

\begin{figure}[!h]
  \centering
  \includegraphics[width=0.5\linewidth]{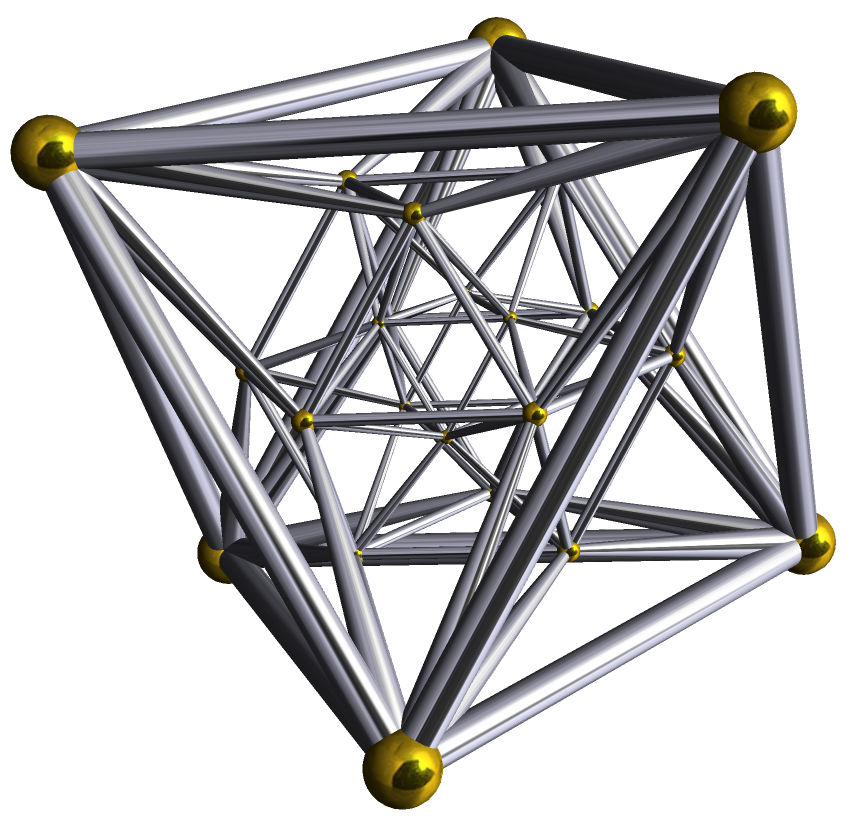}
\caption{The root system of $\mathfrak{so}(8)$ can be identified with a 24-cell. Each vertex corresponds to a fermion flavor.}
\label{fig:24-cell}
\end{figure}

Since the root system of $\mathfrak{so}(8)$ is four-dimensional, it is not as easy to visualize as the root system of $\mathfrak{su}(3)$. It is given by the \textit{24-cell} -- a self-dual four-dimensional polytope with 24 vertices and 24 facets which are (three dimensional) octahedra. See Fig. \ref{fig:24-cell} for an attempt at illustration.

\subsection{Interaction within a generation}

\begin{figure}[!t]
  \centering
  \includegraphics[width=0.8\linewidth]{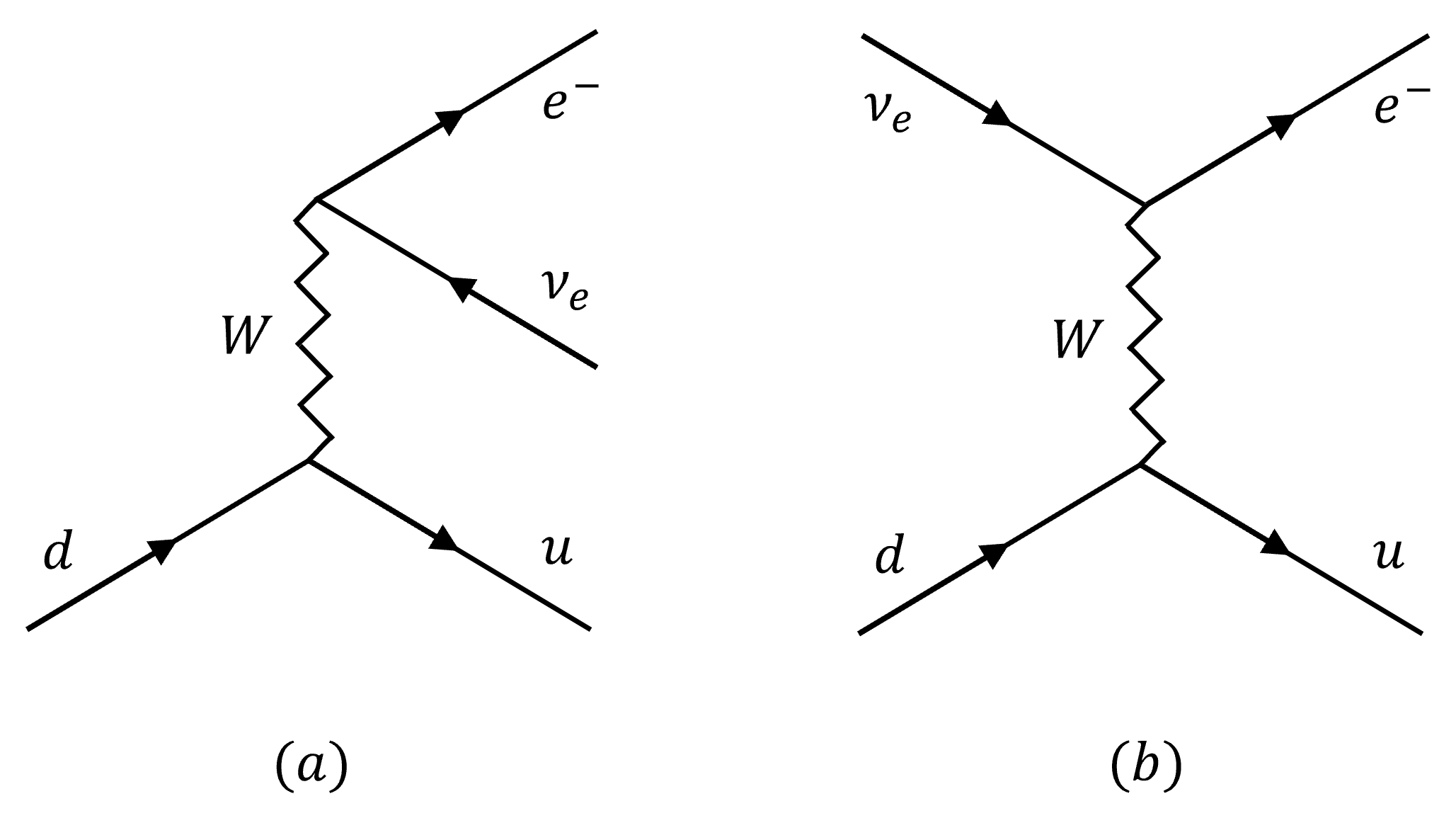}
\caption{(a) The decay of a neutron without spectator quarks. \ \ \ (b) An equivalent but more symmetric interaction.}
\label{fig:neutronfeynman}
\end{figure}

An example of a single-generational interaction is the decay of the neutron, illustrated by the Feyman diagram in Fig. \ref{fig:neutronfeynman}(a). (Here we follow the convention in \cite{Griffiths} that an arrow backwards in time and no bar on the letter indicates an anti-particle.). Removing any spectator quarks, it is given by

\begin{equation}
d \overset{W}{\longrightarrow} u + e^- + \bar{\nu}_e
\label{eq:neutrondecay}
\end{equation}

\noindent If we substitute the letters with their corresponding roots, we will get
\begin{equation}
\underbrace{(1, -1, 0, 0)}_{d} = \underbrace{(1, 1, 0, 0)}_{u} + \underbrace{(0, -1, -1, 0)}_{e^-} + \underbrace{(0, -1, 1, 0)}_{\bar{\nu}_e}
\label{eq:neutrondecayflavors}
\end{equation}
or written in quaternionic notation
$$1 - i = (1 + i) + (-i - j) + (-i + j).$$
The change of ``$\rightarrow$'' in Eq. \ref{eq:neutrondecay} into ``$=$'' in Eq. \ref{eq:neutrondecayflavors} is made on purpose -- it is easily seen that the sum of flavors on the right hand side is actually equal to the flavor on the left hand side. One way to formulate this observation is to say that \textit{flavor is (additively) conserved}. We make the somewhat bold proposal that this is true in general:
\\
\begin{axiom}[Flavor axiom, first version]
In a single-generational interaction, flavor is (additively) conserved.
\end{axiom}
\mbox{}\\
\noindent Also, every single entry of a flavor is (additively) conserved. These entries will be given a special name.
\\
\begin{definition}
Let $(x_0, x_1, x_2, x_3)$ be a fermion flavor, i.e. a root of  $\mathfrak{so}(8)$. The numbers $x_0$, $x_1$, $x_2$ and $x_3$ are called \textbf{fundamental quantum numbers}, or \textbf{flavor numbers}. The number $x_0$ will also be called the \textbf{quark number} of the corresponding particle.
\end{definition}
\mbox{}\\
\noindent Taking a glimpse at Table \ref{table:flavors}, the term ``quark number'' for $x_0$ is quite reasonable: for quarks $x_0 = 1$, for anti-quarks $x_0 = -1$, and for leptons $x_0 = 0$. One more look at the same table explains the indexing: quarks in the $n$th generation have $x_n = \pm 1$ and $x_m = 0$ for $1 \leq m \neq n \leq 3$. 

Up to this point we have only considered the fermions in an interaction, but the $W$ boson also has a very important role to play -- it will \textit{act} on the 24-cell, thereby moving flavor vertices around. This will be more explicit in cross-generational interactions, but it is useful to first understand the concept in the simpler single-generational situation. Since flavor is conserved, we will for this purpose consider the more symmetric interaction in Fig. \ref{fig:neutronfeynman}(b). Here an electron neutrino interacts with a $d$ quark, and the result is an electron and a $u$ quark. We get the corresponding flavor equation by adding an electron neutrino to both sides of Eq. \ref{eq:neutrondecayflavors}:
\begin{equation}
\label{eq:neutronsymmetric}
\underbrace{(1, -1, 0, 0)}_{d} + \underbrace{(0, 1, -1, 0)}_{\nu_e} = \underbrace{(1, 1, 0, 0)}_{u} + \underbrace{(0, -1, -1, 0)}_{e^-}
\end{equation}
This can be understood in the following way: while a $W^+$ increases the quark charge by 1, transforming $d$ into $u$, a $W^-$ decreases the lepton charge by 1, transforming $\nu_e$ into $e^-$. To see how this can be accomplished by some simple quaternionic arithmetic, let us write Eq. \ref{eq:neutronsymmetric} in quaternionic notation:
$$\underbrace{(1 - i)}_{d} + \underbrace{(i - j)}_{\nu_e} = \underbrace{(1 + i)}_{u} + \underbrace{(-i - j)}_{e^-}$$
Thus we want $W^+(1 - i) = 1 + i$ and $W^-(i - j) = -i - j$, which will work if $W^+ = i$ and $W^- = -k$ and the action is given by multiplication. We see that $W^+$ and $W^-$ are elements of the group $Q_8$. 

So far, so good. But this was only the first generation. What happens if we consider a similar interaction within the second or third generation? Let us take for example
$$c + \bar{\nu}_\mu \overset{W}{\longrightarrow} s + \mu^+,$$
or in terms of flavors
$$\underbrace{(1 + j)}_{c} + \underbrace{(-j + k)}_{\bar{\nu}_\mu} = \underbrace{(1 - j)}_{s} + \underbrace{(j + k)}_{\mu^+}.$$
Here $W^-$ instead acts on the $c$ quark and turns it into an $s$ quark. The action is $s = -j\cdot c$, so for the second generation $W^- = -j$ on quarks, while $W^+ = i$ on leptons. For an interaction within the third generation, we would get $W^\pm = \pm k$ (acting on quarks). Thus, we need a pair $W^\pm$ for each generation:
$$W^\pm_1 = \pm (i, k),\ \ \ \ \ W^\pm_2 = \pm (j, i),\ \ \ \ \ W^\pm_3 = \pm (k, j),$$
where the first entry acts on quarks and the second entry on leptons. The six unit quaternions in the first (or second) entries are elements of $Q_8$. So what is really happening in a single-generational electroweak interaction is that the group $Q_8$ acts on the 24-cell given by the roots of $\mathfrak{so}(8)$.

\subsection{Cross-generational interactions}\label{crossgen}

Let us now consider a cross-generational weak interaction. Removing the spectator quarks from the decays $\Lambda \rightarrow p + \pi^-$ and $\Omega^- \rightarrow  \Lambda + K^-$, both will result in the interaction
$$s \overset{W}{\longrightarrow} u + \bar{u} + d.$$
\mbox{}\\
Writing out the flavors, $u$ and $\bar{u}$ will cancel, and we get
$$(1, 0, -1, 0) \neq (1, -1, 0, 0).$$
So flavor is not conserved over generations. Where does it go wrong? Was the Flavor axiom a bad idea after all, or can we modify it in a reasonable way in order to include cross-generational interactions?

The answer to the first question is that elements of $Q_8$ can only change flavor \textit{within} a generation. To fix this problem, we need to extend $Q_8$ to a bigger symmetry which also acts across the generations. The correct choice turns out to be the \textit{binary tetrahedral group}\footnote{This group has been suggested as a fundamental symmetry of nature in \cite{Wilson21}. In the same paper, the idea of using quaternions for flavors is touched upon, but with a different set up than ours.}:
$$T_{24} = Z_3 \ltimes Q_8$$
This group, which has 24 elements, is the semidirect product of $Z_3$ with $Q_8$, where the generator $\omega = -\frac{1}{2}(1 + i + j + k)$ of $Z_3$ acts on $Q_8$ by conjugation. This action will cyclically rotate $i$, $j$ and $k$. $T_{24}$ can be described as the following group of quaternions:
$$T_{24} \cong \{\pm 1, \pm i, \pm j, \pm k, \frac{1}{2}(\pm 1 \pm i \pm j \pm k)\}$$
with all different sign combinations in the last entry. It is quite remarkable that this is \textit{also} the vertices of (the dual of) a 24-cell.

So which element $W \in T_{24}$ describes the cross-generational flavor change $s \rightarrow d$, i.e. which element maps $1 - j$ to $1 - i$? Since $\omega: i \mapsto j$ we must have $W = \omega^{-1}$.

\begin{figure}[!t]
  \centering
  \includegraphics[width=0.8\linewidth]{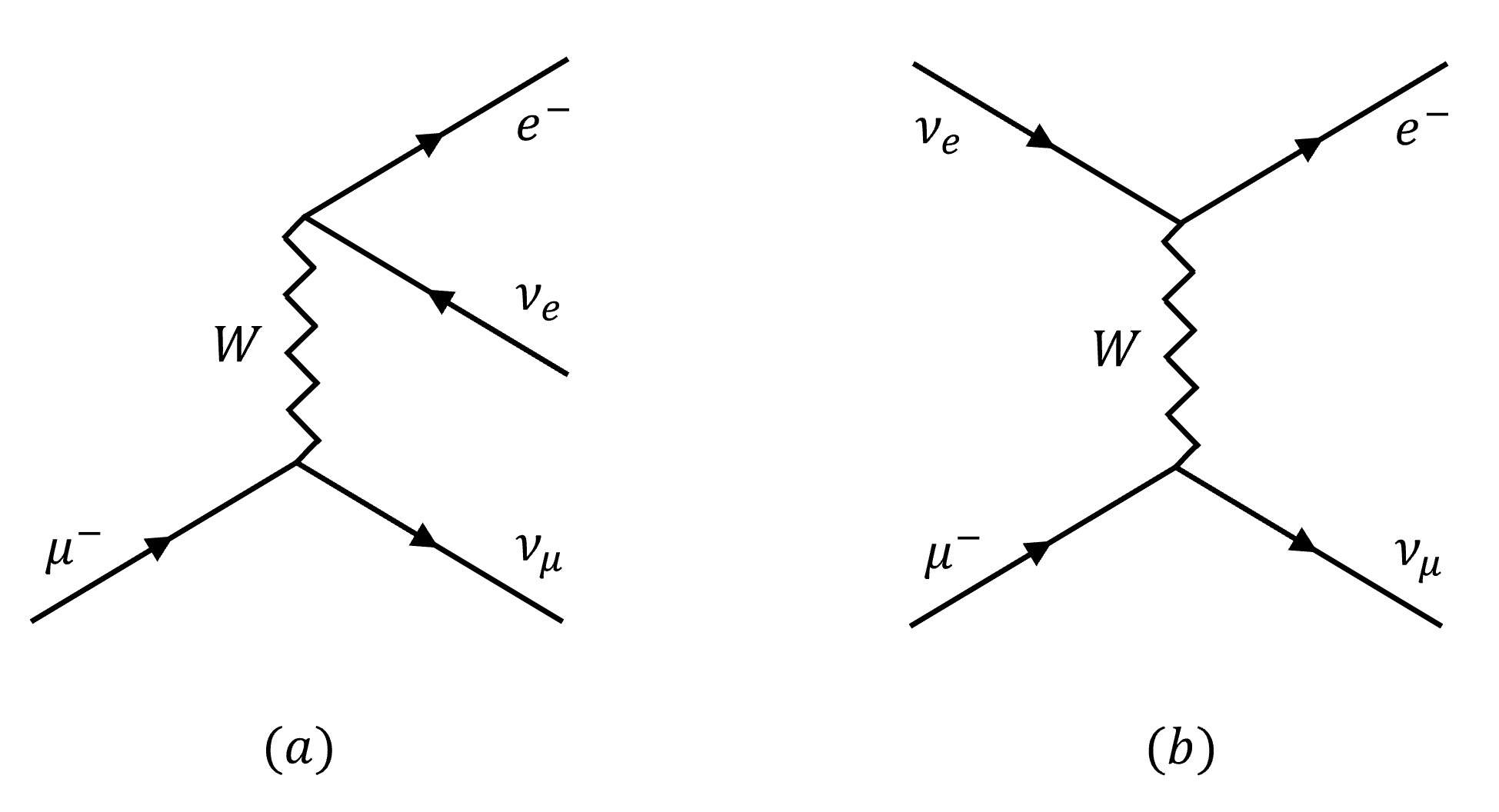}
\caption{(a) Decay of a muon.\ \ \ (b) An equivalent but more symmetric diagram.}
\label{fig:muondecay}
\end{figure}

Let us look at another cross-generational interaction, namely the decay of a muon (see Fig. \ref{fig:muondecay}). The more symmetric version of the diagram corresponds to the flavor equation
$$\underbrace{(0, 0, -1, -1)}_{\mu^-} + \underbrace{(0, 1, -1, 0)}_{\nu_e} \overset{Z_3}{=} \underbrace{(0, -1, -1, 0)}_{e^-} + \underbrace{(0, 0, 1, -1)}_{\nu_\mu},$$
where $Z_3$ above the equality sign indicates ``equality modulo the action of $Z_3$.'' Here $Z_3$ acts by changing generation of both fermions:
$$\omega^{-1}(\mu^-) = e^-\ \ \ \ \mbox{and}\ \ \ \ \omega(\nu_e) = \nu_\mu$$
So when two particles interact, in some sense they follow a law of action and reaction, but on a quantum level: $\omega \in T_{24}$ acts on one particle, while at the same time $\omega^{-1}$ acts on the other.

We are now in a position to extend the Flavor axiom to all electroweak interactions:

\begin{axiom}[Flavor axiom, final version]
In an electroweak interaction flavor is transformed by the action of the binary tetrahedral group.
\end{axiom}

\noindent Or in other words: \textit{Electroweak interaction is the action of the binary tetrahedral group on the $D_4$ root system, i.e. the action of the 24-cell on itself}.

\subsection{Definition of electric charge}

In the Standard Model the (generalized) Gell-Mann--Nishijima formula \cite{Nishijima, Gell-Mann} expresses electric charge $Q$ in terms of six quantum numbers:
$$Q = I_3 + \frac{1}{2}(\mathcal{B} + S + C + B + T),$$
where $I_3$ is the third component of isospin, $\mathcal{B}$ is the baryon number, $S$ is strangeness, $C$ is charmness, $B$ is bottomness, and $T$ is topness. Our ambition is to substitute these (and all other electroweak) quantum numbers with the four fundamental quantum numbers $x_0, x_1, x_2, x_3$, and so it would be desirable to find a similar formula for $Q$ in terms of these four numbers. To this end, let us make the ansatz
$$Q(x_0, x_1, x_2, x_3) = \sum_{n=0}^3 a_nx_n,$$
where we need to find the coefficients $a_0, a_1, a_2$ and $a_3$. Considering the first generation of fermions we get the simultaneous equations
$$\left\{\begin{array}{l}
Q(u) = Q(1, 1, 0, 0) = a_0 + a_1 = \frac{2}{3} \\
Q(d) = Q(1, -1, 0, 0) = a_0 - a_1 = -\frac{1}{3} \\
Q(e^+) = Q(0, 1, 1, 0) = a_1 + a_2 = 1\\
Q(\nu_e) = Q(0, 1, -1, 0) = a_1 - a_2 = 0\\
\end{array}\right.$$
which have the unique solution $a_0 = \frac{1}{6}$ and $a_1 = a_2 = \frac{1}{2}$. Setting up the corresponding equations for the second and third generation will, for symmetry reasons, give the same values and in addition $a_3 =\frac{1}{2}$. Hence we arrive at
\begin{equation}
\label{eq:electriccharge}
Q = \frac{1}{2}\left(\frac{x_0}{3} + x_1 + x_2 + x_3\right).
\end{equation}

\noindent This relationship can be used to \textit{define} electric charge.
\\
\begin{definition}
The \textbf{electric charge} $Q(x)$ of a particle of flavor $x = (x_0, x_1, x_2, x_3)$ is given by Eq. \ref{eq:electriccharge}.
\end{definition}
\mbox{}\\
\noindent Note that for leptons $x_0 = 0$, so we get the simpler formula
$$Q = \frac{1}{2}(x_1 + x_2 + x_3).$$
Thus, for a lepton the electric charge is given by the mean value of the two non-zero entries in the flavor.

\begin{figure}[!t]
  \centering
  \includegraphics[width=\linewidth]{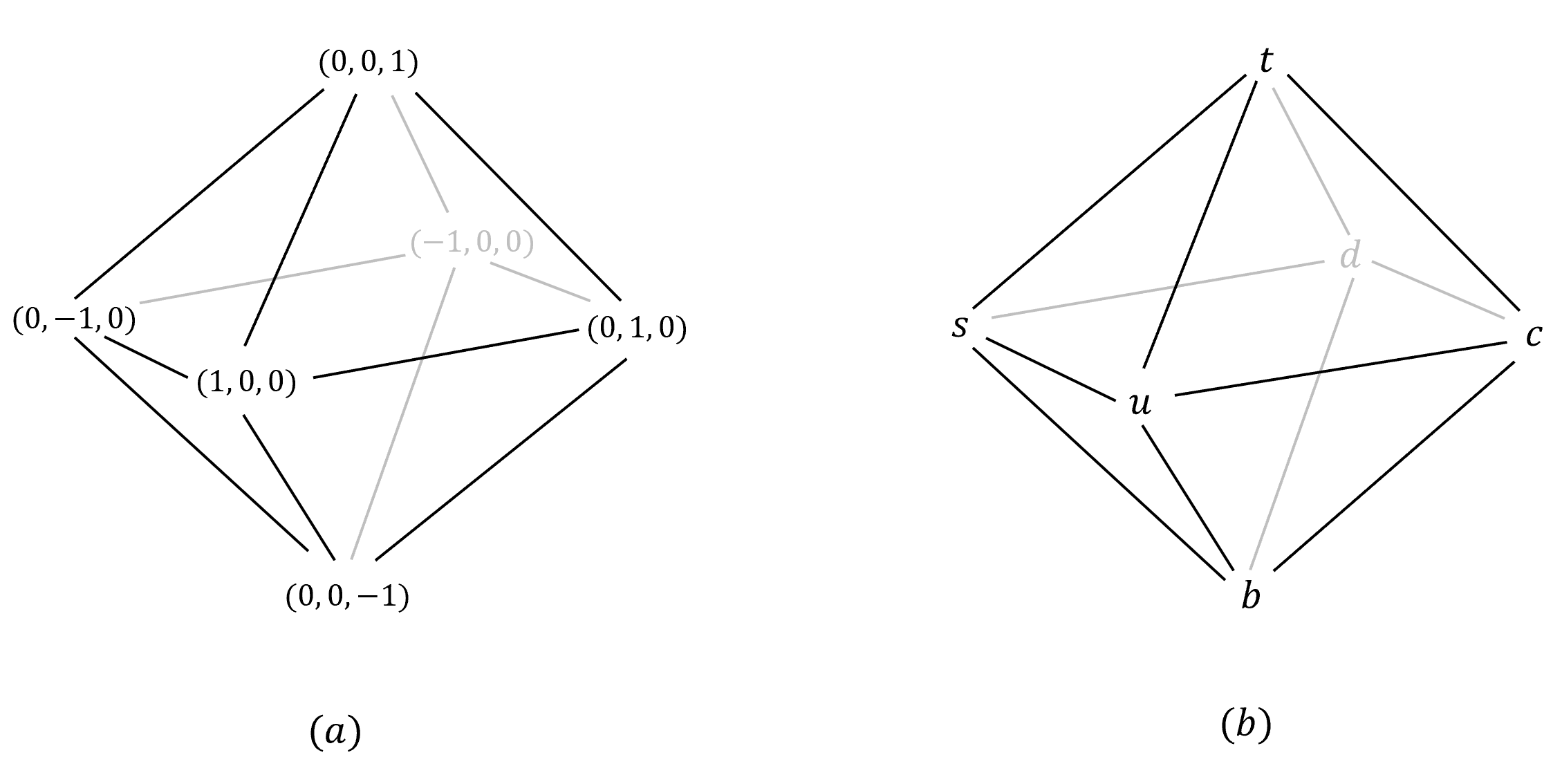}
\caption{(a) The slice $x_0=1$ showing $(x_1, x_2, x_3)$.\ \ \ (b) The quark slice with flavors.}
\label{fig:quarkslice}
\end{figure}

\begin{figure}[!t]
  \centering
  \includegraphics[width=\linewidth]{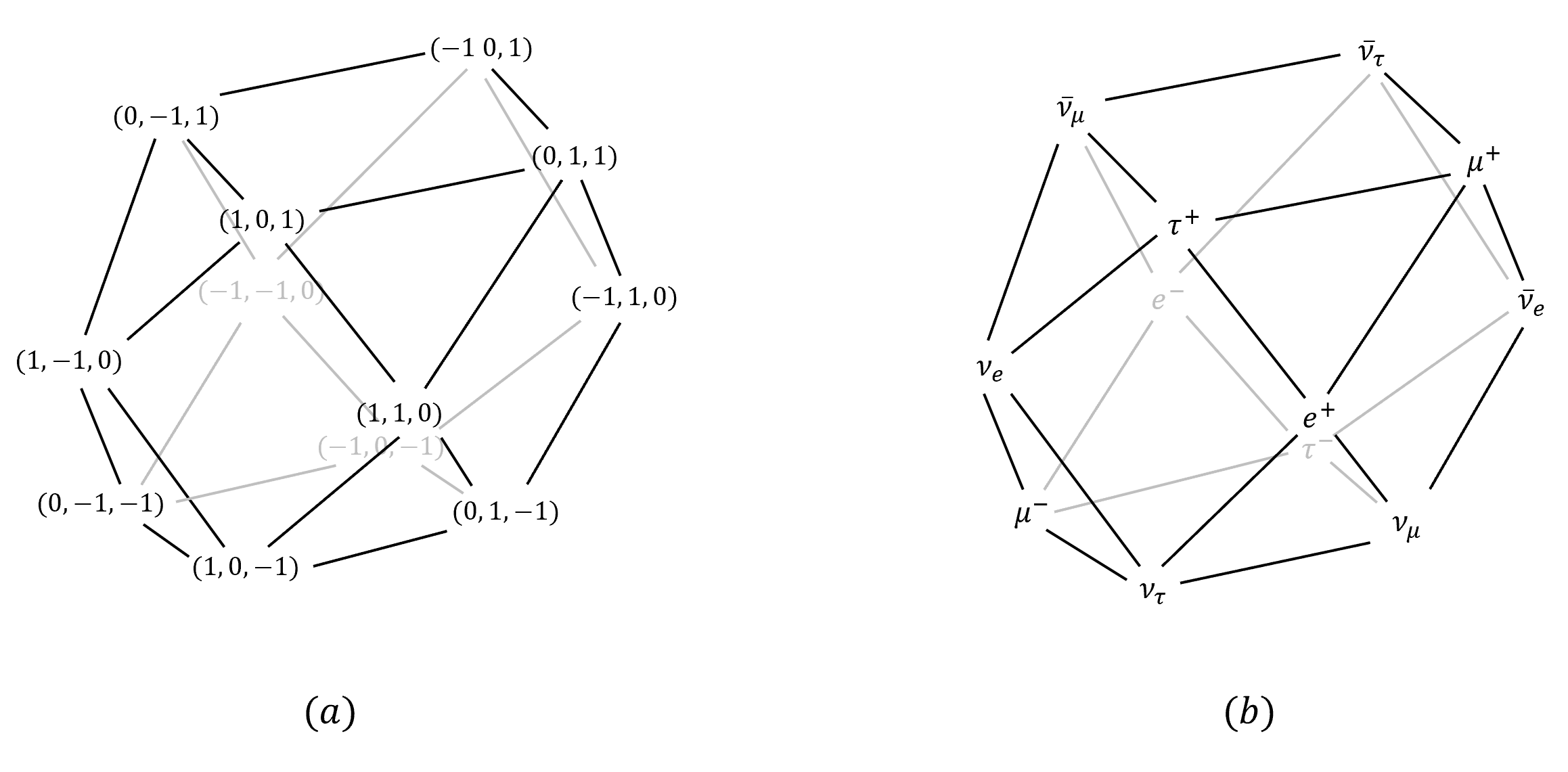}
\caption{(a) The slice $x_0=0$ showing $(x_1, x_2, x_3)$.\ \ \ (b) The lepton slice with lepton types.}
\label{fig:leptonslice}
\end{figure}
In Fig. \ref{fig:quarkslice}, the quarks are shown -- this is the octahedral slice $x_0 = 1$ of the 24-cell. There is an identical octahedron for the anti-quarks ($x_0 = -1$). Fig. \ref{fig:leptonslice} shows the lepton slice $x_0 = 0$: a cuboctahedron containing all the leptons. Note that these quark number slices can be further divided into charge slices. For example, in the lepton slice, there are two triangular charge slices $Q = 1$ and $Q= -1$ for the charged leptons, while the six neutrinos are found in the hexagonal slice $Q = 0$. 

The expression for electric charge in the definition is apparently very similar to the Gell-Mann--Nishijima formula, but has some advantages:
\begin{itemize}
\item It expresses $Q$ with four quantum numbers instead of six.
\item It is also valid for leptons, not only quarks.
\item It treats all quarks on the same footing.
\end{itemize}
The relations between the fundamental quantum numbers and the other quantum numbers in the Gell-Mann--Nishijima formula are
$$(I_3, S, C, B, T) = \frac{1}{2}(x_1, x_2 - 1, x_2 + 1, x_3 - 1, x_3 + 1).$$

\subsection{Conserved quantities}

Since a baryon consists of three quarks, quark number $x_0$ is related to baryon number $\mathcal{B}$ by $x_0 = 3\mathcal{B}$. Now the action of $Z_3$ only permutes $i$, $j$ and $k$, so for this action quark/baryon number is invariant. For the action of $Q_8$, the situation is more complicated. The element $1 \in Q_8$ obviously does not change anything, while $-1 \in Q_8$ actually turns each of the quantum numbers $x_n$, $n = 0, 1, 2, 3$, hence also electric charge, into its negative.\footnote{For a possible physical interpretation of the elements 1 and $-1$ in $Q_8$, see the last section of this article.} Finally, the remaining elements $\pm i, \pm j, \pm k \in Q_8$ will leave $x_0$ unchanged since $\pm i(1 \mp i) = 1 \pm i$ and similarly for the second and third generations.

From the Flavor axiom and the definition of electric charge it is possible to prove the following result.
\\
\begin{theorem}
In all electroweak interactions where quark number is conserved, electric charge is also conserved.
\end{theorem}

\begin{proof}
Let us first prove additivity of $Q$. To this end, let $x = (x_0, x_1, x_2, x_3)$ and $y = (y_0, y_1, y_2, y_3)$ be two flavors so that
$$Q(x) = \frac{1}{2}\left(\frac{x_0}{3} + x_1 + x_2 + x_3\right)$$
and
$$\quad Q(y) = \frac{1}{2}\left(\frac{y_0}{3} + y_1 + y_2 + y_3\right).$$
We then have
\begin{eqnarray*}
Q(x + y) & = & Q(x_0 + y_0, x_1 + y_1, x_2 + y_2, x_3 + y_3) \\
             & = & \frac{1}{2}\left(\frac{x_0 + y_0}{3} + (x_1 + y_1) + (x_2 + y_2) + (x_3 + y_3)\right) \\
             & = & \frac{1}{2}\left(\frac{x_0}{3} + x_1 + x_2 + x_3\right) + \frac{1}{2}\left(\frac{y_0}{3} + y_1 + y_2 + y_3\right) \\
             & = & Q(x) + Q(y).
\end{eqnarray*}
Next, let us note that $Q$ is invariant under the action of $Z_3$, i.e. $Q(\omega^k.x) = Q(x)$ for $k\in\{0, 1, 2\}$. This is true because the action of $Z_3$ does not affect $x_0$, while only permuting the numbers $x_1$, $x_2$ and $x_3$, hence leaving their sum $x_1 + x_2 + x_3$ invariant.

Now consider a single-generational interaction
$$x + y \rightarrow z + w$$
where $Z_3$ acts trivially. By assumption, the weak boson does not correspond to $-1 \in Q_8$ and so we have $x + y = z + w$. It follows from additivity that
$$Q(x) + Q(y) = Q(x + y) = Q(z + w) = Q(z) + Q(w),$$
i.e. the total charge before and after the interaction is the same.

For a cross generational interaction where $Z_3$ acts non-trivially, i.e. $x + y \overset{Z_3}{=} z + w$ with $z = \omega^k.x$ and $w = \omega^{-k}.y$, $k \in \{1, 2\}$, it follows from the $Z_3$ invariance that $Q(x) + Q(y) = Q(z) + Q(w)$.
\end{proof}

\section{Conclusion}\label{sec13}

In this paper we have suggested a classification of elementary fermions in terms of the root system of $\mathfrak{so}(8)$, which can be identified with the 24-cell. At the same time another copy of the (self-dual) 24-cell has been assigned to the electroweak bosons, which are considered elements of the binary tetrahedral group, $T_{24} = Z_3 \ltimes Q_8$. When both fermions and bosons are written in quaternionic notation, the 24-cell will act on itself by some simple arithmetic in $\mathbb{H}$, hence giving the electroweak interactions. By introducing four fundamental quantum numbers, one of which is the quark number, we have given a unifying description of quarks and leptons, as well as a simple definition of electric charge. The elements $\pm i, \pm j, \pm k \in Q_8$ are identified with the charged bosons $W^\pm$ acting on different generations. It remains to give an interpretation of the elements $\pm 1 \in Q_8$. Naturally enough, 1 can be seen to correspond to the electromagnetic boson $\gamma$, i.e. the photon, and the neutral weak boson $Z^0$ -- none of these changes any quantum numbers. So what about $-1$? As mentioned above, when this element acts on a flavor (i.e. a root of $\mathfrak{so}(8)$), it will change all quantum numbers into their negatives, hence transforming a particle to its corresponding anti-particle. While everything else we have discussed so far has been an attempt to explain existing physics in simpler and more profound terms, this last element of $Q_8$ calls for some new physics beyond the Standard Model, e.g. dark matter. In some recent papers \cite{Abbas, Nomura}, dark matter and the flavor problem in connection with discrete symmetries have been adressed. In our situation, a boson corresponding to $-1$ will work as a ``charge conjugator,'' but no such particle has been observed. One might speculate that such charge conjugators contribute to the halo of dark matter surrounding galaxies, thus shielding them off from anti-matter. This might explain the abundance of matter compared to the scarce existence of anti-matter -- the anti-matter is actually there in intergalactic space, but it will never reach us inside the halo. If this is the correct interpretation or if there is some other, possibly more plausible, explanation for this missing piece of the $T_{24}$ symmetry of nature, remains to be seen.

\section{Acknowledgements}

The author wish to thank Bo Sundborg for useful discussions. The author also want to thank Robert Webb -- the 24-cell in Fig. \ref{fig:24-cell} was drawn with his Stella software: http://www.software3d.com/Stella.php


\begin{thebibliography}{99}


\bibitem{Case} Case, K.M., Karplus, R., Yang, C.N., \textit{Strange Particles and the Conservation of Isotopic Spin}, Radiation Laboratory, University of California, Berkeley, 1955.
\bibitem{Wilczek} Wilczek, F., Zee, Z., \textit{Discrete flavor symmetries and a formula for the Cabibbo angle}, Physical Letters, vol. 70B, Issue 4, 1977.
\bibitem{Frampton94} Frampton, P.H., Kephart, T.W., \textit{Simple Non-Abelian Finite Flavor Groups and Fermion Masses},  	Int. J. Mod. Phys. A 10:4689-4704, 1995.
\bibitem{Aranda} Aranda, A., Carone, C.D., Lebed, R.F., \textit{Maximal Neutrino Mixing from a Minimal Flavor Symmetry}, Phys. Rev. D 62:016009, 2000.
\bibitem{Feruglio} Feruglio, F., Hagedorn, C., Lin, Y., Merlo, L., \textit{Tri-bimaximal Neutrino Mixing and Quark Masses from a Discrete Flavour Symmetry}, Nucl. Phys. B 775:120-142, 2007; Erratum-ibid. 836:127-128, 2010.
\bibitem{Eby} Eby, D.A., Frampton, P.H., \textit{Nonzero $\theta_{13}$ signals nonmaximal atmospheric neutrino mixing}, Phys. Rev. D 86 (2012) 117304.
\bibitem{King} King, S. F., Luhn, C., \textit{Neutrino Mass and Mixing with Discrete Symmetry}, Rept. Prog. Phys. 76, 056201 (2013).
\bibitem{Frampton23} Frampton, P.H., Corianò, C., Santorelli, P., \textit{Atmospheric Neutrino Octant from Flavour Symmetry},  arXiv:2305.10463v2 [hep-ph], 17 May 2023.
\bibitem{Wilson20} Wilson, R.A., \textit{A group-theorist's perspective on symmetry groups in physics},  arXiv:2009.14613v5 [math.GR], 20 December 2020.
\bibitem{Wilson211} Wilson, R.A., \textit{Options for a finite group model of quantum mechanics},  arXiv:2104.10165v5 [math.GR], 7 July 2021.
\bibitem{Wilson212} Wilson, R.A., \textit{Subgroups of Clifford algebras}, Adv. App. Clifford Algebras 31 (2021), 59.
\bibitem{Wilson21} Wilson, R.A., \textit{Finite symmetry groups in physics},  arXiv:2102.02817v5 [math.GR], 8 November 2023.
\bibitem{Baez} Baez, J., \textit{Two of my favorite numbers: 8 and 24}. Online lecture, Harvard University, 24 September, 2023. https://math.ucr.edu/home/baez/8\_and\_24/



\bibitem{Huerta} Baez, J., Huerta, J., \textit{The Algebra of Grand Unified Theories}, Bull. Am. Math. Soc. 47:483-552, 2010.
\bibitem{Yang} Yang, C.N., Mills, R., \textit{Conservation of Isotopic Spin and Isotopic Gauge Invariance}, Phys. Rev. 96, 191 (1954).
\bibitem{Ross} Ross, G.G., \textit{Grand Unified Theories}, The Benjamin/Cummings Publishing Company, Inc., 1985.
\bibitem{Smith} Smith, F.D., \textit{Spin(8) Gauge Field Theory}, Int. J. Theor. Phys., vol. 25, Issue 4, 1986. 
\bibitem{Pati} Pati, J.C., Salam, A., \textit{Lepton Number as the Fourth Color}, Phys. Rev. D, 10:275–289, 1974.


\bibitem{Conway12} Conway, A.W., \textit{The quaternionic form of relativity}, Phil. Mag. 24 (1912), 208.
\bibitem{Conway48} Conway, A.W., \textit{Quaternions and quantum mechanics}, Ponteacrè Acad. Sci. Acta 12 (1948), 204–277.
\bibitem{Dirac} Dirac, P.A.M., \textit{Applications of quaternions to Lorentz transformations}, Proc. Roy. Irish Acad. A 50 (1945), 261–270.
\bibitem{Lambek} Lambek, J., \textit{If Hamilton Had Prevailed: Quaternions in Physics}, The Mathematical Intelligencer vol. 17, no. 4, 1995.
\bibitem{Gursey} Günaydin, M., Gürsey, F., \textit{Quark structure and octonions}, J. Math. Phys. 14, 1651–1667 (1973).
\bibitem{Furey} Furey N., Hughes, M.J., \textit{Division algebraic symmetry breaking}, Physics Letters B 831 (2022). 
\bibitem{Furey2} Furey, N., \textit{An Algebraic Roadmap of Particle Theories, Part I: General construction}, arXiv:2312.12377 [hep-ph].
\bibitem{Furey3} Furey, N., \textit{An Algebraic Roadmap of Particle Theories, Part II: Theoretical checkpoints}, arXiv:2312.12799 [hep-ph].
\bibitem{Furey4} Furey, N., \textit{An Algebraic Roadmap of Particle Theories, Part III: Intersections}, arXiv:2312.14207 [hep-ph].



\bibitem{Gell-Mann61} Gell-Mann, M., \textit{The Eightfold Way: A Theory of Strong Interaction Symmetry}, Synchroton Laboratory, California Institute of Technology, Pasadena, 1961.
\bibitem{Gell-Mann} Gell-Mann, M., \textit{The interpretation of the new particles as displaced charge multiplets}, Il Nuovo Cimento, vol 4 (1956).
\bibitem{Nishijima} Nishijima, K., \textit{Charge Independence Theory of V Particles}, Progress of Theoretical Physics, Volume 13, Issue 3 (1955).
\bibitem{Yokota} Yokota, I. \textit{Exceptional Lie Groups}. arXiv:0902.0431v1 [math.DG] 3 Feb 2009.
\bibitem{Griffiths} Griffiths, D. \textit{Introduction to Elementary Particles}, 2nd edition, Wiley-VCH, 2008.


\bibitem{Abbas} Abbas, G., Adhikari, R., Chun, E.J., \textit{Flavonic dark matter}, Phys. Rev. D 108, 115035, 2023.
\bibitem{Nomura} Nomura, T., Shimizu, Y., Takahashi, T., \textit{Flavino dark matter in a non-Abelian discrete flavor model}, JHEP09 (2024) 036.

\end{thebibliography}
\end{document}